\documentclass[11pt, reqno]{article}

\usepackage{jheppub}
\usepackage{amssymb}
\usepackage{amsmath}
\usepackage{amsthm}
\usepackage[usenames,dvipsnames]{xcolor}
\usepackage{epsfig}
\usepackage{dcolumn}
\usepackage{tikz}
\usetikzlibrary{shapes.geometric, arrows}
\usepackage{upgreek}
\usepackage{setspace}
\usepackage{enumitem}
\usepackage{array,multirow,bigdelim,arydshln}
\usepackage{appendix}
\usepackage{xparse}
\usepackage{stmaryrd}
\usepackage{graphicx}

\usepackage[nottoc]{tocbibind}

\usepackage[utf8]{inputenc}
\hypersetup{
	colorlinks,
	urlcolor=Maroon,
	linkcolor=Maroon,
	citecolor=Maroon
}

\NewDocumentCommand{\binomial}{omm}
 {%
  \genfrac(){0pt}{}{#2}{#3}%
  \IfValueT{#1}{_{\!#1}}%
 }
\NewDocumentCommand{\eulerian}{omm}
 {%
  \genfrac<>{0pt}{}{#2}{#3}%
  \IfValueT{#1}{_{\!#1}}%
 }

\def \s {\sigma}

    \usepackage{latexsym}
    \usepackage{tikz}

\theoremstyle{plain}

\theoremstyle{definition}

\title{Notes on Biadjoint Amplitudes, ${\rm Trop}\,G(3,7)$ and $X(3,7)$ Scattering Equations}

\author[a]{Freddy Cachazo}\emailAdd{fcachazo@pitp.ca}
\author[a,b,c]{and Jairo M. Rojas}\emailAdd{jmrojash@pucp.edu.pe}

\affiliation[a]{Perimeter Institute for Theoretical Physics, Waterloo, ON N2L 2Y5, Canada}
\affiliation[b]{Department of Physics $\&$ Astronomy, University of Waterloo, Waterloo, ON N2L 3G1, Canada}
\affiliation[c]{ICTP South American Institute for Fundamental Research\\
Instituto de F\'{i}sica Te\'{o}rica, UNESP-Universidade Estadual Paulista\\ R. Dr. Bento T. Ferraz 271, Bl. II, S\~{a}o Paulo 01140-070, SP, Brazil}

\abstract{In these notes we use the recently found relation between facets of tropical Grassmannians and generalizations of Feynman diagrams to compute all ``biadjoint amplitudes" for $n=7$ and $k=3$. We also study scattering equations on $X(3,7)$, the configuration space of seven points on $\mathbb{CP}^2$. We prove that the number of solutions is $1272$ in a two-step process. In the first step we obtain $1162$ explicit solutions to high precision using near-soft kinematics. In the second step we compute the matrix of $360\times 360$ biadjoint amplitudes obtained by using the facets of ${\rm Trop}\, G(3,7)$, subtract the result from using the $1162$ solutions and compute the rank of the resulting matrix. The rank turns out to be $110$, which proves that the number of solutions in addition to the $1162$ explicit ones is exactly $110$.
}

\begin{document}
\maketitle
\addtocontents{toc}{\protect\setcounter{tocdepth}{1}}
\def \tr {\nonumber\\}
\def \la  {\langle}
\def \ra {\rangle}
\def\hset{\texttt{h}}
\def\gset{\texttt{g}}
\def\sset{\texttt{s}}
\def \be {\begin{equation}}
\def \ee {\end{equation}}
\def \ba {\begin{eqnarray}}
\def \ea {\end{eqnarray}}
\def \k {\kappa}
\def \h {\hbar}
\def \r {\rho}
\def \l {\lambda}
\def \be {\begin{equation}}
\def \en {\end{equation}}
\def \bes {\begin{eqnarray}}
\def \ens {\end{eqnarray}}
\def \red {\color{Maroon}}
\def \pt {{\rm PT}}
\def \s {\sigma} % Is there no problem with line 117?
\def \ls {{\rm LS}}
\def \ma {\Upsilon}
\def \s {\textsf{s}}
\def \t {\textsf{t}}
\def \R {\textsf{R}}
\def \W {\textsf{W}}
\def \U {\textsf{U}}
\def \e {\textsf{e}}

\numberwithin{equation}{section}

\section{Introduction}

Very recently Early, Guevara, Mizera, and one of the authors introduced and studied a natural generalization of the scattering equations, which connect the space of Mandelstam invariants to that of points on ${\mathbb{CP}^1}$ \cite{Fairlie:1972zz,Fairlie:2008dg,Cachazo:2013gna,Cachazo:2013hca}, to higher dimensional projective spaces $\mathbb{CP}^{k-1}$ \cite{Cachazo:2019ngv}. In this work we only study $k=3$. The equations are obtained by computing the critical points of a potential function
\be\label{gen}
{\cal S} \equiv \sum_{1\leq a_1<a_2<a_3\leq n} \s_{a_1a_2a_3}\log\,( a_1,a_2,a_3).
\ee
Here $\s_{a_1a_2a_3}$ are a generalization of Mandelstam invariants while $( a_1,a_2,a_3)$ can be thought of as Pl\"ucker coordinates on $G(3,n)$. The configuration space of $n$ points on $\mathbb{CP}^{2}$ is obtained by modding out by a torus action $\mathbb{C}^*$ on each of the points, i.e., $X(3,n):= G(3,n)/(\mathbb{C}^*)^n$ \cite{sekiguchi1997w}.

The kinematic invariants satisfy
\be
\s_{aab} = 0, \qquad \sum_{b,c}\s_{abc} = 0 \qquad \forall\, a.
\ee
These are the $k=3$ analogs of masslessness and momentum conservation conditions. These conditions guarantee that the potential function is invariant under the torus action and therefore one can choose inhomogeneous coordinates for points on $\mathbb{CP}^2$ to be $(x_i,y_i)$. The Pl\"ucker coordinates are then replaced by
\be
|abc| := {\rm det}\left(
                    \begin{array}{ccc}
                      1 & 1 & 1 \\
                      x_a & x_b & x_c \\
                      y_a & y_b & y_c \\
                    \end{array}
                  \right).
\ee

Having a higher $k$ version of the scattering equations, one can immediately generalize the Cachazo-He-Yuan (CHY) formula for biadjoint amplitudes \cite{Cachazo:2013iea}. Such amplitudes can then be expanded in terms of higher $k$ Feynman diagrams as defined by facets of the tropical Grassmanian \cite{Cachazo:2019ngv}. Even though a direct quantum field theory derivation of such generalized Feynman diagrams is not yet known, it is important to mention that the known physically relevant case of $k=2$ implies via the duality $X(2,n)\sim X(n-2,n)$ that $k=n-2$ is also physically relevant. It is therefore very natural to complete the spectrum of possibilities between the two ends, i.e., for $2<k<n-2$. At the very least one can expect that a good understanding of such cases would deepen that of the $k=2$ and $k=n-2$ cases.

Given two orderings $\alpha$ and $\beta$ one defines \cite{Cachazo:2019ngv}
\be\label{biad}
m_n^{(3)}(\alpha|\beta) := \frac{1}{{\rm vol}(SL(3,\mathbb{C}))}\int \prod_{a=1}^n dx_a\,dy_a \sideset{}{'}\prod_{a=1}^{n}\delta\left(\frac{\partial {\cal S}}{\partial x_a}\right)\delta\left(\frac{\partial {\cal S}}{\partial y_a}\right)  {\rm PT}^{(3)}(\alpha){\rm PT}^{(3)}(\beta)
\ee
where $k=3$ Parke-Taylor functions \cite{Parke:1986gb} are naturally generalized to
\be
{\rm PT}^{(3)}(1,2,\ldots , n) := \frac{1}{|123|\, |234|\,\cdots \, |n12|}.
\ee
CHY integrals are localized to the solutions of the scattering equations \cite{Cachazo:2013hca}. For $k=2$ it is known that the number of solutions is $(n-3)!$. For $k=3$ it is known that there are $2$ and $26$ solutions for $n=5$ and $n=6$ respectively \cite{Cachazo:2019ngv}. Also in \cite{Cachazo:2019ngv} two lower bounds for the number of solutions for $n=7$ were obtained using soft limit arguments. The stronger of the two is $1152$. In this note we prove that the number of solutions is $1272$. This number is consistent with a computation done by T. Lam using uniform matriods over finite fields which also reproduces all other known results \cite{PC}.

In \cite{Cachazo:2019ngv} the authors provided evidence that CHY integrals defined on $X(k,n)$ have a deep and surprising connection to the tropical Grassmannian ${\rm Trop}\,G(k,n)$. The connection for $k=2$ is fairly direct. It maps all possible $2^{n-1}-n-1$ propagators to all possible vertices (or rays) of ${\rm Trop}\,G(2,n)$. In addition, there is also a bijection between every one of the $(2n-5)!!$ tree-level Feynman diagrams in a cubic scalar theory and facets of ${\rm Trop}\,G(2,n)$. 

In \cite{Cachazo:2019ngv} the case of $n=6$ and $k=3$ was studied in detail. In this work we extend the same analysis to $n=7$, $k=3$ biadjoint amplitudes and ${\rm Trop}\,G(3,7)$. Luckily, the structure of ${\rm Trop}\,G(3,7)$ has been carefully studied by Herrmann, Jensen, Joswig, and Sturmfels \cite{HJJS} and we make use of their results to carry out all our Feynman diagram computations which lead to the explicit form of all $360\times 360$ biadjoint amplitudes $m_7^{(3)}(\alpha|\beta)$.

The strategy for determining the number of solutions to the $X(3,7)$ scattering equations is the following. We start with the dual version, i.e., the $X(4,7)$ scattering equations near a soft limit. The strict soft limit gives equations on $X(4,6)\backsimeq X(2,6)$ and produces $6$ solutions. Each solution is then used to find a polynomial of degree $192$ from the equations for the soft particle. This produces $6\times 192=1152$ seeds of solutions from which a numerical search for the true solutions can be launched. In the kinematics we use this search returned $1162$ solutions which we obtained with more than $2000$ digits of precision.

Having the explicit form of all $360\times 360$ biadjoint amplitudes $m_7^{(3)}(\alpha|\beta)$ and their corresponding approximation $m_7^{(3):{\rm approx}}(\alpha|\beta)$ from using \eqref{biad} with $1162$ solutions, we compute the $360\times 360$ matrix of differences  $\Delta_7^{(3)}(\alpha|\beta ) := m_7^{(3)}(\alpha|\beta)-m_7^{(3):{\rm approx}}(\alpha|\beta)$ and its rank. The result turns out to be $110$. This is the number of missing solutions which gives the result stated above of $1272$ solutions.

This paper is organized as follows: Section 2 contains a review of the proposal connecting tropical Grassmannians and generalized Feynman diagrams with special emphasis on ${\rm Trop}\, G(3,7)$. After identifying the list of planar poles a prescription is given for computing $m_7^{(3)}(\alpha |\beta)$. In section 3 we present a computation of $1162$ explicit solutions to the $X(3,7)$ scattering equations. Comparing with the results of the previous section it is concluded that there are still some solutions missing. In section 4 we apply a technique that allows the determination of the number of missing solutions without explicitly finding them. We end in section 5 with discussion and some future directions.

\section{Biadjoint Amplitudes from the Tropical Grassmannian $G(3,7)$}

Grassmannian spaces have appeared in a variety of scattering amplitudes constructions and some of them include the Grassmannian formulation of ${\cal N}=4$ super Yang-Mills amplitudes in twistor space, momentum space and momentum twistor space (see \cite{ArkaniHamed:2012nw} for a review). So far the Grassmannians $G(k,n)$ have been standard in the sense of being defined as algebraic varieties over $\mathbb{R}$ or $\mathbb{C}$. Very recently, Grassmannians defined as tropical varieties have made a surprising appearance in the physics of scattering amplitudes \cite{Cachazo:2019ngv}. Tropical geometry has also naturally appeared in the study of string theory amplitudes \cite{Tourkine:2013rda} as well as in mirror symmetry, e.g. see \cite{mirror}.

We are only interested in the $k=3$ case. The tropical Grassmannian $G(3,n)$ can be thought of as polyhedral complex in $\mathbb{R}^{n \choose 3}$. In fact there are several versions of it. Here we follow the original description given by Speyer and Sturmfels in their study of ${\rm trop}\,G(3,6)$ \cite{SS} but adapted to the work of Herrmann, Jensen, Joswig, and Sturmfels \cite{HJJS} where the analysis of ${\rm trop}\,G(3,7)$ is done.  

In order to describe the object of interest let us introduce $\e_{i_1,i_2,i_3}$ to be a basis of unit vectors in $\mathbb{R}^{n \choose 3}$. One has to mod out by a map
\be
\phi: (a_1,a_2,\ldots ,a_n) \to \sum_{i_1,i_2,i_3} (a_{i_1}+a_{i_2}+a_{i_3})\e_{i_1,i_2,i_3}
\ee
which turns out to be related to momentum conservation in its physical application.                                                                                                  
We are interested in the image of the tropical Grassmannian in $\mathbb{R}^{n\choose 3}/{\rm image}\,\phi$. The corresponding object is also a polyhedral complex which abusing the notation we also denote by ${\rm Trop}\,G(3,n)$.

Here we use and check the proposal made in \cite{Cachazo:2019ngv} relating rays of ${\rm Trop}\,G(3,n)$ with ``Feynman propagators" and Gr\"obner cones (or facets if intersected with a unit sphere) with ``Feynman diagrams". Both generalized objects compute ``amplitudes" which are formally obtained using the $k=3$ CHY formulation, in particular, those of the biadjoint theory \eqref{biad}.

In order to explain the map we specialize to the case of interest, i.e., ${\rm Trop}\,G(3,7)$. The case ${\rm Trop}\,G(3,6)$ was discussed in \cite{Cachazo:2019ngv}. In ${\rm Trop}\,G(3,7)$ there are $721$ rays. Each ray can be expressed in terms of the basis of vectors as
\be
r^{(I)} = \sum_{i_1<i_2<i_3}c^{(I)}_{i_1i_2i_3}\e_{i_1i_2i_3}.
\ee
The list of all $721$ rays can be found in:  

\url{www.uni-math.gwdg.de/jensen/Research/G3_7/grassmann3_7.html}

The list is labeled ``R-vector" and each entry is a row with $35$ elements. For example, the row labeled as $90$ reads
\be
(1,1,1,1,1,0,0,\cdots , 0)
\ee
and represents the ray $\e_{123}+\e_{124}+\e_{125}+\e_{126}+\e_{127}$.

This combination seems strange at first by once we mod out by $\phi$ this vector is also equivalent to 
\be
\e_{345}+\e_{346}+\e_{347}+\e_{356}+\e_{357}+\e_{367}+\e_{456}+\e_{457}+\e_{467}+\e_{567}
\ee
which resembles a kinematic invariant. 

Following \cite{Cachazo:2019ngv}, for each ray we associate a propagator according to the map
\be
r^{(I)} \to \frac{1}{\sum_{i_1<i_2<i_3}c^{(I)}_{i_1i_2i_3}\s_{i_1i_2i_3}}.
\ee

The table of $721$ rays is separated into six classes. The translation of all classes of rays into physical language was given in \cite{Cachazo:2019ngv} and reproduced here for the reader's convenience
\be
    \centering
    \begin{tabular}{|c|c|}
    \hline
       [0,34]  & $1/\s_{123}$ \\ \hline
       [35,69] & $1/\t_{1234}$ \\ \hline
       [70,90] & $1/\t_{12345}$ \\ \hline
        [91,300] & $1/(\t_{12345}+\s_{456}+\s_{457})$ \\ \hline
        [301,615] & $1/(\t_{1234}+\t_{1256}+\s_{127})$ \\ \hline
        [616,720] & $1/(\t_{12345}+t_{34567}+\t_{56712})$ \\
        \hline
    \end{tabular}\label{tab:classes7}\nonumber
\ee
Note that the row $90$ given above corresponds to 
\be
\frac{1}{\t_{34567}} := \frac{1}{\sum_{3\leq i<j<k\leq 7}\s_{ijk}}.
\ee

The first three classes are familiar from the $k=2$ perspective. The fourth class is the $n=7$ analog of the new class present in $n=6$ and denoted by $R$ in \cite{Cachazo:2019ngv}. The last two classes are new as they first appear for $n=7$. We introduce the notation
\be
\R_{abcdefg} :=
  \t_{abcde} + \s_{def} + \s_{deg}, \,\, \W_{abcdefg} :=
  \t_{abcd} + \t_{abef} + \s_{abg}, \,\,  \U_{abcdefg} :=
 \t_{abcde} + \t_{cdefg} + \t_{efgab}.
\ee

The next step in the translation between the mathematical object ${\rm Trop}\,G(3,7)$ and generalized amplitudes is the connection between cones or facets of ${\rm Trop}\,G(3,7)$ and Feynman diagrams. 

Facets are classified into $125$ types and given in the same webpage provided above in a table labeled ``Gr\"obner cone" and in the column called ``Rays". Each facet is determined by a set of six rays therefore identified by a list of six positions in the list of $721$ rays.

Here we present the first ten facets translated into physics as an illustration of the map:
\begin{equation*}
\begin{array}{|c|c|}
\hline
 1 & \left\{\s_{567},\s_{347},\s_{246},\s_{235},\s_{145},\s_{136}\right\} \\
 \hline
 2  & \left\{\s_{567},\s_{347},\s_{246},\s_{235},\s_{145},\W_{\text{1523674}}\right\} \\ \hline
 3  & \left\{\s_{567},\s_{347},\s_{246},\s_{145},\s_{123},\W_{\text{1523674}}\right\} \\ \hline
 4  & \left\{\s_{567},\s_{347},\s_{246},\s_{235},\t_{1235},\W_{\text{1235467}}\right\} \\ \hline
 5  & \left\{\s_{567},\s_{347},\s_{246},\s_{135},\t_{1235},\W_{\text{1235467}}\right\} \\ \hline
 6 & \left\{\s_{567},\s_{347},\s_{246},\s_{145},\R_{\text{2346715}},\W_{\text{1523674}}\right\} \\ \hline
 7  & \left\{\s_{567},\s_{347},\s_{246},\s_{135},\t_{1235},\R_{\text{2564713}}\right\} \\ \hline
 8  & \left\{\s_{567},\s_{347},\s_{246},\s_{145},\R_{\text{2346715}},\W_{\text{3415267}}\right\} \\ \hline
 9 & \left\{\s_{567},\s_{347},\s_{246},\s_{135},\R_{\text{2564713}},\W_{\text{5613247}}\right\} \\ \hline
 10  & \left\{\s_{567},\s_{347},\s_{246},\s_{135},\R_{\text{2564713}},\R_{\text{1345726}}\right\} \\ \hline
  \end{array}
 \end{equation*}{}

The $k=3$ Feynman diagram associated with the first facet is then
\be
\frac{1}{\s_{567}\,\s_{347}\,\s_{246}\,\s_{235}\,\s_{145}\,\s_{136}}.
\ee
Clearly, this particular Feynman diagram does not contribute to biadjoint amplitudes as the collection of all its poles is not simultaneously compatible with any planar ordering.

\subsection{Computation of Biadjoint Amplitudes Using ${\rm Trop}\,G(3,7)$}

Now we are ready to explain the computation of biadjoint amplitudes $m_{7}^{(3)}(\alpha|\beta)$ also following the proposal made in \cite{Cachazo:2019ngv}.

The first step is to determine every possible pole which is compatible with a given planar ordering (or planar). Consider first the canonical ordering $\mathbb{I}:=1234567$:
\be
    L(\mathbb{I})=\{\s_{123},\t_{1234},\t_{12345},\R_{1234567},\R_{3217654},\U_{1234567},\W_{1234675},\ldots\}
\ee
where the ellipses indicate all cyclic permutations\footnote{Here we follow the definition of kinematic invariants given in \cite{Cachazo:2019ngv} and it turns out that the indices in $\W_{1234675}$ for $\mathbb{I}:=1234567$ do not appear in the natural order.}. The set $L(\mathbb{I})$ has $49$ elements. This is surprising at first as one would expect that restricting to a given planar ordering forces the restriction from ${\rm Trop}\,G(3,7)$ to its positive part which is known to have $42$ vertices or rays \cite{SW}. The reason this is consistent is that $\U_{1234567}$  and its cyclic versions are all spurious as we explain below.  

%\begin{equation}
%\begin{gathered}
%    \qty{\mathsf{t}_{1234},\mathsf{t}_{3456},\mathsf{t}_{5612}, R_{12,34,56}}, \\
%    \qty{\mathsf{t}_{1234},\mathsf{t}_{3456},\mathsf{t}_{5612}, R_{12,56,34}}.
%    \end{gathered}
%\end{equation}

A Feynman diagram is then said to be planar if all its six poles are planar. We searched every possible facet of ${\rm Trop}\,G(3,7)$ for planar diagrams and found $1064$. These are all simplices and might not be true facets of ${\rm Trop}\,G(3,7)$ as some of them can combine to form more complicated polytopes. Luckily, our planarity condition is enough to pick the correct combinations of simplices without inputting any further information.

Using these $1064$ facets one computes their corresponding $1064$ Feynman diagrams and collect them in a list $J(\mathbb{I})$. The most complicated of the biadjoint amplitudes is then:
\be
m_{7}^{(3)}(\mathbb{I} | \mathbb{I})=\sum_{\Upsilon\in J(\mathbb{I})}\Upsilon.
\ee
In an ancillary \texttt{Mathematica} notebook titled {\it Tropical Grassmannian} we provide all $1064$ Feynman diagrams for the canonical order and the definition of all kinematic invariants involved in them.

In order to compute any other amplitude $m_{7}^{(3)}(\alpha|\beta)$ we apply a permutation of labels to $J(\mathbb{I})$ to produce the set of Feynman diagrams compatible with the ordering of interest, $J(\alpha)$ and $J(\beta)$, and finally sum over the intersection of the sets:
\begin{equation}
    m_{7}^{(3)}(\alpha|\beta)=(-1)^{\omega(\alpha,\beta)}\sum_{\Upsilon\in J(\alpha)\cap J(\beta)}\Upsilon.
\end{equation}
As it is familiar from the $k=2$ biadjoint scalar theory, each sum over Feynman diagrams has to be multiplied by an overall sign function $(-1)^{\omega(\alpha,\beta)}$. The function can be computed following the way the color decomposition leads to $m_{7}^{(2)}(\alpha|\beta)$. At this point, a color structure has not been identified for $k=3$ and therefore the sign function is computed with a combination of the CHY formula \eqref{biad} and soft limit consistency. The results are presented in the  \texttt{Mathematica} notebook mentioned above.

We close this section by providing an example:
%\begin{equation*}
%    m_3^{(7)}(1234567|1342756)=\frac{1}{\s_{234} \s_{567} \t_{1234} %\t_{12347} \t_{12567} \t_{1567}},
%\end{equation*}{}
\begin{eqnarray*}
    m_3^{(7)}(\mathbb{I}|1236457) & = &\frac{1}{\R_{6543217} \s_{456} \t_{12367} \t_{1237} \t_{23456} \t_{3456}}+\frac{1}{\R_{4567123} \s_{127} \s_{456} \t_{12367} \t_{1237} \t_{4567}}\\&+&\frac{1}{\R_{4567123} \s_{456} \t_{12367} \t_{1237} \t_{14567} \t_{4567}}+\frac{1}{\R_{4567123}\R_{6543217} \s_{456} \t_{12367} \t_{1237} \t_{14567}}\\&+&\frac{1}{\R_{6543217} \s_{123} \s_{456} \t_{12367} \t_{1237} \t_{14567}}+\frac{1}{\R_{4567123} \R_{6543217} \s_{456} \t_{12367} \t_{1237} \t_{23456}}\\&+&\frac{1}{\R_{4567123} \s_{127} \s_{456} \t_{12367} \t_{1237} \t_{23456}}+\frac{1}{\R_{6543217} \s_{123} \s_{456} \t_{12367} \t_{1237} \t_{3456}}\\&+&\frac{1}{\s_{127} \s_{456} \t_{12367} \t_{1237} \t_{34567} \t_{4567}}+\frac{1}{\s_{127} \s_{456} \t_{12367} \t_{1237} \t_{23456} \t_{3456}}\\&+&\frac{1}{\s_{123} \s_{456} \t_{12367} \t_{1237}  \t_{14567} \t_{4567}}+\frac{1}{\s_{123} \s_{456} \t_{12367} \t_{1237} \t_{34567} \t_{4567}}\\&+&\frac{1}{\s_{123} \s_{456} \t_{12367} \t_{1237} \t_{3456} \t_{34567}}+\frac{1}{\s_{127} \s_{456} \t_{12367} \t_{1237} \t_{3456} \t_{34567}}+T_{\U},
\end{eqnarray*}
where $T_{\U}$ contains all terms with a $\U$-pole and is given by
\begin{eqnarray*}
T_{\U} & := & \frac{1}{\R_{4567123} \R_{6543217} \t_{23456} \U_{2345671} \s_{456}\t_{12367}}+\frac{1}{\R_{4567123} \R_{6543217} \s_{456} \t_{12367} \t_{14567} \U_{2345671}}\\ &+&\frac{1}{\R_{4567123} \R_{6543217} \s_{456} \t_{14567} \t_{23456} \U_{2345671}}.
\end{eqnarray*}

One can check that, as explained above, the three Feynman diagrams containing the pole $\U_{2345671}$ combine to cancel it. The presence of spurious poles in some formulations of amplitudes is familiar \cite{Britto:2004ap,Britto:2005fq,Hodges:2009hk} and it is often useful to make certain properties manifest. In the case at hand, the $\U$ poles make the connection to the facets of ${\rm Trop}\, G(3,7)$ direct.

\section{Computation of $1162$ Solutions to $X(3,7)$ Scattering Equations}

In the standard case of the scattering equations on $X(2,n)$ the number of solutions is easily computed using an induction argument based on a soft limit: One takes all kinematic invariants that involve particle $n$ and scale them to zero using a parameter $\tau$. In the limit $\tau\to 0$ the scattering equations split into those for $n-1$ particles and one equation which determines the position of the $n^{\rm th}$ particle once the other ones are known. This last equation has $n-3$ solutions and therefore one concludes that the number of solutions for $n$ particles is related to that of $n-1$ particles by ${\cal N}_n^{(k=2)} = (n-3){\cal N}_{n-1}^{(k=2)}$. Since ${\cal N}_{3}^{(k=2)}=1$ one concludes that ${\cal N}_{n}^{(k=2)}=(n-3)!$.

In \cite{Cachazo:2019ngv}, the same induction argument was done for $k=3$. It was found that ${\cal N}_6^{(3)} = 13\,{\cal N}_{5}^{(3)}$ and since ${\cal N}_{5}^{(3)}={\cal N}_{5}^{(2)}=2$, then ${\cal N}_6^{(3)} =26$. Also, ${\cal N}_7^{(3):regular} = 42\,{\cal N}_{6}^{(3)}=1092$. Note the superscript $regular$ which indicates that there are solutions which escape the soft limit analysis as they involve singular configurations for the hard particles. One reason it is known that there are missing solutions is that the same analysis done for $k=4$ leads to ${\cal N}_7^{(4):regular} = 192\,{\cal N}_{6}^{(4)}$. Since ${\cal N}_{6}^{(4)}={\cal N}_{6}^{(2)}=6$, then ${\cal N}_7^{(4):regular} =1152$. Using the isomorphism between $X(4,7)$ and $X(3,7)$ it is clear that ${\cal N}_7^{(3)}\geq 1152>{\cal N}_7^{(3):regular}$. In the next section we prove that even this $1152$ is only a lower bound and the number of singular solutions is ${\cal N}_7^{(4):singular} = 120$. Therefore the total number of solutions is ${\cal N}_7^{(4)} =1272$. Using the duality again one concludes that ${\cal N}_7^{(3):singular} = 180$.  

The first step in the proof requires the computation of the largest possible number of solutions to the $X(3,7)$ equations to very high numerical precision. This sounds like a daunting task for even the $X(2,n)$ scattering equations are very challenging to solve for $n=9$ where the number of solutions is only $6!=720$. Our strategy is to use kinematics that is close to a soft limit in order to use the regular solutions as seeds for a numerical search of the true solutions to the system. Given that ${\cal N}_7^{(k=4):regular}$ is larger than ${\cal N}_7^{(k=3):regular}$ we choose to apply the soft limit technique to $X(4,7)$ and then translate to $X(3,7)$ language. 

Before explaining the procedure let us review the map connecting $X(4,7)=G(4,7)/(\mathbb{C}^*)^7$ and $X(3,7)=G(3,7)/(\mathbb{C}^*)^7$. The isomorphism follows from that of the underlying Grassmannians. The map for Grassmannians is standard and starts with a gauge fixed version of a matrix representative of $G(4,7)$
\begin{equation} C = 
	\begin{pmatrix}
	1 & 0 & 0 & 0 & c_{15} & c_{16}   & c_{17}   \\
	0 & 1 & 0 & 0 & c_{25} & c_{26}   & c_{27} \\
	0 & 0 & 1 & 0 & c_{35} & c_{36}   & c_{37} \\
	0 & 0 & 0 & 1 & c_{45} & c_{46}   & c_{47} \\
	\end{pmatrix}.
\end{equation}
The map takes any $4\times 4$ minor of $C$ to a $3\times 3$ minor of 
\begin{equation}
\tilde{C}= \begin{pmatrix}
c_{17} & c_{27} &c_{37} & c_{47} & 0 & 0 & -1 \\
c_{16} & c_{26} &c_{36} & c_{46} & 0 & -1 & 0 \\
c_{15} & c_{25} &c_{35} & c_{45} & -1 & 0 & 0 \\
\end{pmatrix}.
\end{equation}
More explicitly, the map takes $|abcd| = \epsilon^{abcdefg}|efg|$, where the set $\{abcdefg\}=\{1,2,\ldots,7\}$, i.e., repeated indices are not summed, and $\epsilon$ is the completely antisymmetric tensor.

Clearly, since all our formulas are invariant under the torus action and the corresponding $GL(4,\mathbb{C})$ and $GL(3,\mathbb{C})$ redundancies, it is easy to go from one space to the other.

\subsection{Strict Soft Limit: Finding Seeds}

We start by using the torus action to choose the following complete gauge fixing on $X(4,7)$:
\begin{equation}
	\begin{pmatrix}
	1 & 0 & 0 & 0 & 1 & 1   & 1   \\
	0 & 1 & 0 & 0 & 1 & x_6 & x_7 \\
	0 & 0 & 1 & 0 & 1 & y_6 & y_7 \\
	0 & 0 & 0 & 1 & 1 & z_6 & z_7 \\
	\end{pmatrix}.
\end{equation}

Next we construct a one parameter family of kinematic invariants which interpolates between generic $k=4$, $n=6$ kinematics and generic $k=4$, $n=7$ kinematics. In the strict soft limit $\tau\to 0$ and assuming that we are only dealing with regular solutions, particle seven decouples from the equations for the first six particles and the scattering equations reduce to those of $k=2$, $n=6$ for which it is easy to find all six solutions explicitly. Let us denote them by $\{x_6^{(i):{\rm soft}},y_6^{(i):{\rm soft}},z_6^{(i):{\rm soft}}\}$.

For each of the six solutions particle seven is forced to satisfy its own scattering equations 
\begin{equation}\label{poly7}
\sum_{\substack{1\leq a < b < c \leq 6}}\frac{{\hat \s}_{7abc}(abc)^{(i)}_{xy}}{|7abc|_{(i)}} = 0\,, \,\,
\sum_{\substack{1\leq a < b < c \leq 6}}\frac{{\hat \s}_{7abc}(abc)^{(i)}_{yz}}{|7abc|_{(i)}} = 0 \,\, \text{  and }
\!\sum_{\substack{1\leq a < b < c \leq 6}}\frac{{\hat \s}_{7abc}(abc)^{(i)}_{zx}}{|7abc|_{(i)}} = 0. \ \
\end{equation}
Here $\s_{7abc} =\tau {\hat\s}_{7abc}$ and
\be
(abc)_{xy} := \left|
                      \begin{array}{ccc}
                        1 & 1 & 1 \\
                        x_a & x_b & x_c \\
                        y_a & y_b & y_c \\
                      \end{array}
                    \right|.
\ee
The definition of $(abc)^{(i)}_{yz}$ and $(abc)^{(i)}_{zx}$ is completely analogous. The superscript $(i)$ indicates that all coordinates for particle six are evaluated on the $i^{\rm th}$ solution to the soft equations. 

The three equations \eqref{poly7} for $x_7,y_7,z_7$ turn out to have $192$ solutions. We find all $6\times 192=1152$ solutions in order to use them as initial conditions for a numerical search of solutions to the complete set of equations on kinematics near, but at a finite distance, from the strict soft limit. 

\subsection{Near-Soft Kinematics: Finding True Solutions}

Having computed all $1152$ regular solutions to the strict soft limit we proceed to increase the parameter $\tau$ so that the ratio of kinematic invariants involving only hard particles to that of $\s_{7abc}$ is around $10^4$. This choice establishes a hierarchy but it is otherwise generic. 

Let us denote a set of six independent scattering equations on the near soft kinematics as 
\be
{\cal E}_{\rm near\,soft} := \left\{ \frac{\partial {\cal S}}{\partial x_6},\frac{\partial {\cal S}}{\partial y_6},\frac{\partial {\cal S}}{\partial z_6},\frac{\partial {\cal S}}{\partial x_7},\frac{\partial {\cal S}}{\partial y_7},\frac{\partial {\cal S}}{\partial z_7} \right\}.
\ee
We performed a numerical search for roots of ${\cal E}_{\rm near\,soft}$ near each of the $1152$ solutions. This is easily done using the Mathematica function \texttt{FindRoot}. The function also gives an option for setting the working precision. Using this we found 1142 solutions near the initial conditions and $20$ which came in complex conjugated pairs. Each pair lies near one of the remaining $10$ initial conditions. The total of $1162$ solutions where found with $2000$ digits of precision. 

In an ancillary \texttt{Mathematica} notebook titled {\it 1162 Solutions} we provide the kinematics and gauge fixing used for the computation of the solutions. We also provide all $1162$ solutions with few digits of precision (due to size constraints) but also an implementation of \texttt{FindRoot} with the scattering equations which can be used to boost the precision of any one of the solutions to $2000$ digits. 

We evaluated simple biadjoint scalar amplitudes on the $1162$ solutions and found results that agreed up to $98\%$ with the tropical Grassmannian computation indicating that there are some solutions missing which cannot be obtained by the method.

\section{Missing Solutions: Rank of the Matrix of Differences}

Having found a large number of solutions, even larger than the soft limit lower bound, and checked that there must be some missing solutions, we introduce a method for finding the number of missing solutions without having to explicitly find them. The idea is very simple. 

From the CHY formulation, we know that the amplitude $m_7^{(3)}(\alpha|\beta)$ can be expressed as:
\be\label{one}
m_7^{(3)}(\alpha|\beta)
= \sum_{i=1}^{1162}\frac{1}{{\rm det}{'}\Phi^{(3)}_i}{\rm PT}(\alpha)_i{\rm PT}(\beta)_i+\sum_{i=1163}^{\mathcal{N}^{3}_7}\frac{1}{{\rm det}{'}\Phi^{(3)}_i}{\rm PT}(\alpha)_i{\rm PT}(\beta)_i,
\ee
where ${\rm det}{'}\Phi^{(3)}_i$ is the standard CHY jacobian adapted to $k=3$ while $\mathcal{N}^{(3)}_7$ is the total number of solutions. The sum in the first term of \eqref{one} is only over the known $1162$ solutions. 

\begin{figure}[ht]
	\centering
	\includegraphics[width=0.8\textwidth]{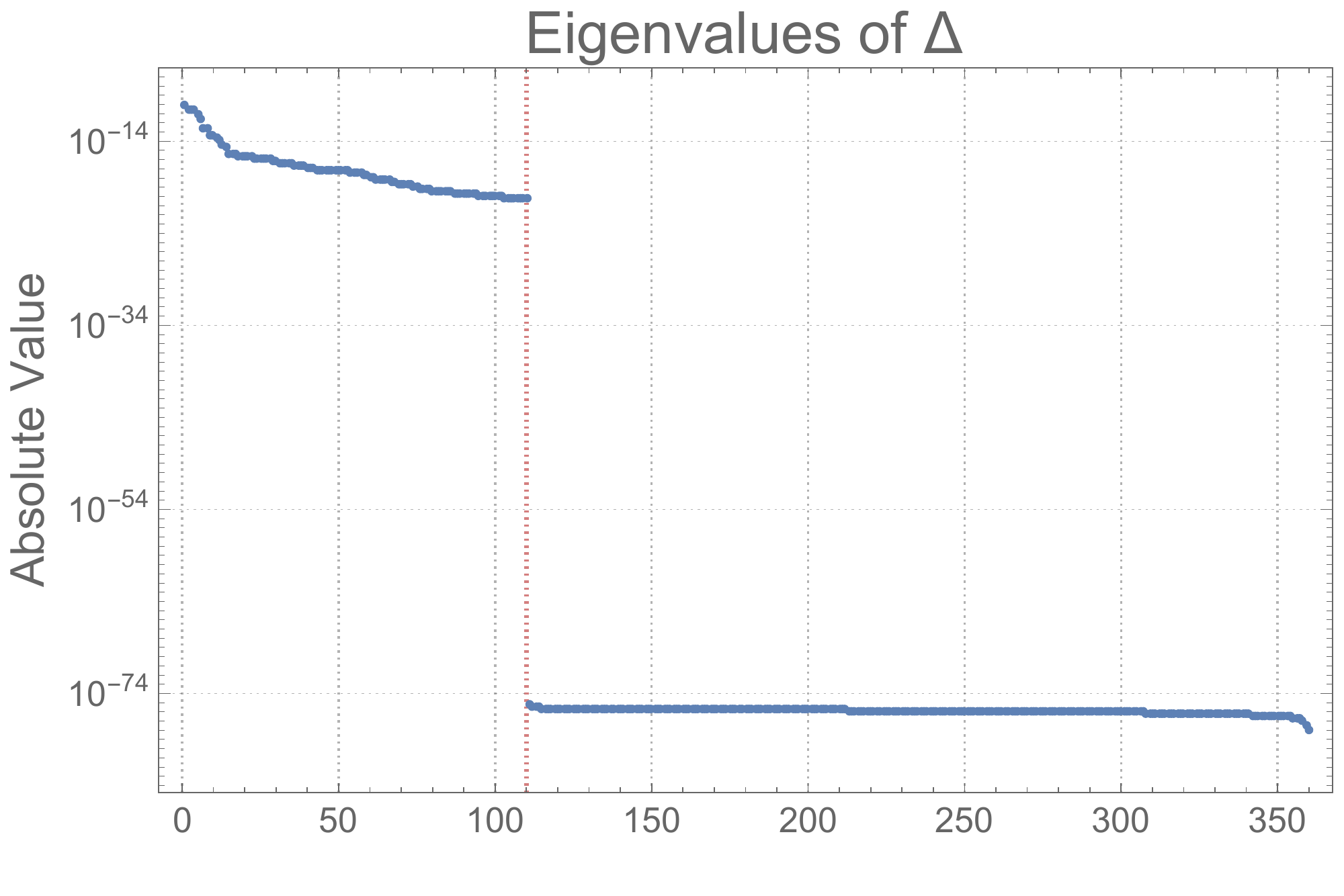}
	\caption{Absolute values of the 360 eigenvalues of the matrix of differences $\Delta_7^{(3)}$. A red vertical line at the location of the 110$^{th}$ eigenvalue is included in order to visualize the separation between non-zero and vanishing eigenvalues (computations were done using only 50 digits of precision).}
	\label{fig:Eigenvalues}
\end{figure}

We now compute the matrix of differences
\be\label{two}
\Delta_7^{(3)}(\alpha|\beta) := m_7^{(3)}(\alpha|\beta)
- \sum_{i=1}^{1162}\frac{1}{{\rm det}{'}\Phi^{(3)}_i}{\rm PT}(\alpha)_i{\rm PT}(\beta)_i.
\ee
In general the number of linearly independent orders is $(n-1)!/2$ which equals $360$ for $n=7$ and therefore $\Delta_7^{(3)}$ is a $360\times 360$ matrix. 

Let us introduce the notation ${\cal M} := \mathcal{N}^{(3)}_7-1162$ for the number of missing solutions. 

According to \eqref{one}, the matrix $\Delta_7^{(3)}$ also admits the structure of the Gram matrix of $360$ vectors $v_\alpha$ living in $\mathbb{C}^{{\cal M}}$, i.e.
\be
\Delta_7^{(3)}(\alpha|\beta) = \sum_{I=1}^{\cal M}v_{\alpha}^I v_{\beta}^I
\ee
where the sum is over the missing solutions (and hence the reason for changing the index from $i$ to $I$) while
\be
v_\alpha^I := \frac{{\rm PT}(\alpha)_I}{\left({\rm det}{'}\Phi^{(3)}_I\right)^{1/2}}.
\ee 

In general the Gram matrix of $m$ vectors in $\mathbb{C}^{{\cal M}}$ has rank equal to ${\cal M}$ if ${\cal M}<m$ and $m$ otherwise. 

The matrix $\Delta_7^{(3)}(\alpha|\beta)$ is also given by \eqref{two} and since $m_7^{(3)}(\alpha|\beta)$ is known from the construction in section 2 using ${\rm Trop}\, G(3,7)$, the matrix $\Delta_7^{(3)}(\alpha|\beta)$ can be evaluated to high numerical precision. 

The computation of the rank of $\Delta_7^{(3)}(\alpha|\beta)$ is most easily done by numerically computing its eigenvalues. The rank, i.e., the number of non-zero eigenvalues, turns out to be $110$ as shown in the table. This means that ${\cal M}=110$ and therefore $\mathcal{N}^{(3)}_7=1272$.

In an ancillary \texttt{Mathematica} notebook titled {\it Number of Solutions} we provide the code for performing the computation of the eigenvalues of the matrix $\Delta_7^{(3)}(\alpha|\beta)$. In order to do so we also provide two data files which contain all elements of the $360\times 360$ matrix of amplitudes $m_7^{(3)}(\alpha|\beta)$ computed using the ${\rm Trop}\, G(3,7)$ titled  {\it mTropmatrix} and another one titled {\it mscEqsolsmatrix} with the $360\times 360$ matrix of approximate amplitudes computed using the $1162$ solutions to high precision. 

As mentioned in the introduction, $\mathcal{N}^{(3)}_7=1272$ also happens to be the result of a computation done using uniform matroids over finite fields\footnote{We are thankful to Song He and Thomas Lam for pointing this out to us.} and it would be very interesting to find out the precise connection to the computation presented in this work.

\section{Discussions}

In this work we computed all biadjoint scattering amplitudes $m_7^{(k=3)}(\alpha|\beta)$ by using the connection between tropical Grassmannians and generalized Feynman diagrams proposed in \cite{Cachazo:2019ngv}. This is done by mapping the vertices of ${\rm Trop}\,G(3,7)$ to all possible poles and then choosing the ones that are compatible with a given planar ordering. Facets of ${\rm Trop}\,G(3,7)$ give rise to Feynman diagrams and we identify planar ones as those that only contain planar poles. 

Of course, a second definition of $m_7^{(k=3)}(\alpha|\beta)$ is provided by the CHY formula based on the scattering equations with Parke-Taylor functions as integrands \cite{Cachazo:2019ngv}. This requires finding all solutions to the scattering equations in order to get values of the amplitudes. We were able to explicitly compute $1162$ solutions to high precision and to prove that the total number is $1272$. 

The proof that the number of solutions in addition to the $1162$ found is $110$ consists in the computation of the rank of the matrix of differences $\Delta_7^{(3)}(\alpha|\beta)$. 
The fact that the rank is $110$, which is less than the dimension of the matrix, implies that the definitions of $m_7^{(k=3)}(\alpha|\beta)$ using the ${\rm Trop}\,G(3,7)$ and using the scattering equations are consistent with each other.

Knowing that the number of singular solutions for $k=4$ and $n=7$ is $120$ and $180$ for $k=3$ and $n=7$ provides a valuable hint about the nature of singular solutions. A very pressing question is to carry out the analytic analysis of the scattering equations by choosing singular coordinates to count the number of such solutions. 

Also worth mentioning is the fact that for the particular kinematic point we studied $1142$ solutions are real as expected very near the soft limit for positive kinematics, as defined in \cite{Cachazo:2019ngv}, while $20$ come in $10$ complex conjugated pairs. This means that some thresholds, i.e., lines where a kinematic invariant changes sign, were crossed. It would be interesting to study the behavior of all kinematic invariants as one moves from the strict soft limit to the near-soft limit configuration and identify these thresholds. Finding them and carefully studying the form of the $20$ complex solutions could provide hints on how to design kinematics so that more solutions could be found using the soft seeds.

\section*{Acknowledgements}

We would like to thank S. He and Y. Zhang for very useful discussions about matroids and for comparing their computation of biadjoint amplitudes using pushforwards to our computation using the tropical Grassmannian. We also thank F. Borges, N. Early, D. Garcia, A. Guevara and S. Mizera for useful discussion. Research at Perimeter Institute is supported in part by the Government of Canada through the Department of Innovation, Science and Economic Development Canada and by the Province of Ontario through the Ministry of Economic Development, Job Creation and Trade. J.M.R. thanks the Abdus Salam International Centre for Theoretical Physics, ICTP-SAIFR/IFT-UNESP and FAPESP grant 2016/01343-7 for partial financial support.

\renewcommand{\thefigure}{\thesection.\arabic{figure}}
\renewcommand{\thetable}{\thesection.\arabic{table}}

%\appendix

%\section{List of Feynman Diagrams in \texorpdfstring{$Trop\ G(3,7)$}{TEXT}}\label{appendix1}

\bibliographystyle{unsrt}
\bibliography{references}

\end{document}